\documentclass[10pt]{IEEEtran}  
\usepackage[pdftex]{hyperref} 

\usepackage{amsmath,amsthm, amssymb, epsfig, verbatim, color, amsfonts} 
\usepackage{graphicx}

\ifx\theorem\undefined
\theoremstyle{plain}
\newtheorem{theorem}{Theorem}

\theoremstyle{definition}
\fi

\def\vr{\mathbf}

\def\Pr{\mathrm{Pr}}
\def\E{\mathbb{E}}

\def\Normal{\mathcal{N}}

\def\Ind{\mathrm{Ind}} 

\def\Var{\mathrm{Var}}

\def\endofproof{\hspace{\stretch{1}}$\Box$}
\def\defeq{\triangleq} 

\def\etal{\textit{et al}}

\def\csiszar{Csisz\'ar~}
\def\korner{K\"orner~}

\newcommand{\tsubs}[1]{{\scriptscriptstyle \mathrm{#1}}}

\newcommand{\arrowabovebelow}[2] {\underset{#2}{\overset{#1}{\textstyle \longrightarrow}}}

\ifx\note\undefined
\newenvironment{note}{\comment}{\endcomment} 
\fi

\newenvironment{inputpath}[1]
{ \let\origtexinput\input \renewcommand{\input}[1]{\origtexinput{#1/##1}}
\let\origtexincludegraphics\includegraphics \renewcommand{\includegraphics}[2][]{\origtexincludegraphics[##1]{#1/##2}} }
{ \let\input\origtexinput \let\includegraphics\origtexincludegraphics}

\allowdisplaybreaks[1]

\title{A simpler derivation of the coding theorem}

\author{Yuval Lomnitz, Meir Feder \\
Tel Aviv University, Dept. of EE-Systems  \\
Email: \{yuvall,meir\}@eng.tau.ac.il}

\begin{document}
\maketitle

\begin{abstract}
A simple proof for the Shannon coding theorem, using only the Markov inequality, is presented. The technique is useful for didactic purposes, since it does not require many preliminaries and the information density and mutual information follow naturally in the proof. It may also be applicable to situations where typicality is not natural.
\end{abstract}

\section{Introduction}\label{sec:intro}
Shannon's channel coding theorem (achievability) for memoryless channels was originally proven based on typicality \cite{Shannon48}, which is formalized in today's textbooks \cite{CoverThomas_InfoTheoryBook} by the asymptotic equipartition property (AEP). The way information theory is introduced in most textbooks and graduate courses, requires one to first get acquainted with the concepts of entropy, mutual information, typicality, etc, before being able to understand this proof. Gallager \cite{GallagerExp65} proposed a different proof leading also to the achievable error exponent. An alternative proof for DMC-s was given by \csiszar and \korner using the method of types \cite[\S IV]{MethodOfTypes}. Less known is Shannon's proof \cite{ShannonCertain57} based on what is nowadays termed information density, which also uses elementary tools.

In recent works the Markov inequality was used to analyze the performance of rateless codes \cite{NavotThesis} and universal decoding schemes \cite{YL_PhdThesis}. The underlying technique is remarkably simple, especially when applied the memoryless channel. In addition to Markov inequality, the proof only uses basic probability laws and the law of large numbers.
This technique may be already known to some, but was never published, so it seems worthwhile to do so.

\section{A proof of the coding theorem}
Random variables are denoted by capital letters and vectors by boldface. When applying a single-letter distribution to a vector it is implicity extended i.i.d., i.e. $P_\tsubs{X}(\vr X_1^n) \defeq \prod_{i=1}^n P_\tsubs{X}(X_i)$.

The Markov inequality simply states that for a non-negative random variable $A$,
\begin{equation}\label{eq:Amarkov_bound}
\Pr \{ A \geq t \} \leq \frac{\E[A]}{t}
\end{equation}
and is easily proven by taking the expected value over the relation $\Ind( A \geq t ) \leq \frac{A}{t}$ (where $\Ind(\cdot)$ denotes an indicator function).

As in the standard proof, the code is a random code where each letter of each codeword is drawn i.i.d. with the distribution $P_\tsubs{X}(\vr X)$. The standard claim that the existence of deterministic capacity achieving codes results from the existence of random codes is applied. After seeing the channel output vector $\vr Y$, the receiver applies maximum likelihood decoding and chooses the codeword $\vr X$ which maximizes $P_\tsubs{Y|X}(\vr Y | \vr X)$ (breaking ties arbitrarily). Note that the decoding metric $P_\tsubs{Y|X}(\vr Y | \vr X)$ does not depend on the specific code chosen.

Now, fix the transmitted and the received words $\vr X, \vr Y$ (respectively) and ask what is the pairwise error probability over the ensemble where the other codeword $\vr X_m$ $m=1,\ldots, 2^{nR}-1$ is independent of $\vr X,\vr Y$ and distributed $P_\tsubs{X}(\cdot)$. Denote by $E_m$ the event that the codeword $\vr X_m$ attains a higher a-posteriori probability, i.e. that $P_\tsubs{Y|X}(\vr Y | \vr X_m) \geq P_\tsubs{Y|X}(\vr Y | \vr X)$. Then
\begin{equation}\begin{split}\label{eq:61}
\Pr \left\{ E_m \big| \vr X,\vr Y \right\}
&=
\Pr \left\{ P_\tsubs{Y|X}(\vr Y | \vr X_m) \geq P_\tsubs{Y|X}(\vr Y | \vr X) \big| \vr X,\vr Y \right\}
\\& \stackrel{\text{Markov}}{\leq}
\frac{ \E  \left[ P_\tsubs{Y|X}(\vr Y | \vr X_m) \big| \vr X,\vr Y \right]}{P_\tsubs{Y|X}(\vr Y | \vr X) }
\\& =
\frac{ \displaystyle \sum_{\vr x_m \in \mathcal{X}^n} P_\tsubs{Y|X}(\vr Y | \vr x_m) P_{X} (\vr x_m) }{P_\tsubs{Y|X}(\vr Y | \vr X)}
\\& =
\frac{ P_\tsubs{Y}(\vr Y)}{P_\tsubs{Y|X}(\vr Y | \vr X)}
.
\end{split}\end{equation}
By the union bound, the probability of error conditioned on $\vr X, \vr Y$ is bounded as:
\begin{equation}\begin{split}\label{eq:72}
P_{e|x,y}
& \leq
\Pr \left\{ \bigcup_{i=1}^{2^{nR}-1} E_m  \Big| \vr X,\vr Y \right\}
\\& \leq
2^{nR} \cdot \Pr \left\{ E_m \big| \vr X,\vr Y \right\}
\\ & \leq
2^{nR} \cdot \frac{ P_\tsubs{Y}(\vr Y)}{P_\tsubs{Y|X}(\vr Y | \vr X)}
\end{split}\end{equation}
Next, the behavior of this conditional error probability $P_{e|x,y}$ is analyzed for the memoryless channel. By the law of large numbers:
\begin{equation}\begin{split}\label{eq:82}
\frac{1}{n} \log \frac{ P_\tsubs{Y}(\vr Y)}{P_\tsubs{Y|X}(\vr Y | \vr X)}
&=
\frac{1}{n} \sum_{i=1}^n \log \frac{ P_\tsubs{Y}(Y_i)}{P_\tsubs{Y|X}(Y_i | X_i)}
\\& \arrowabovebelow{\text{in Prob. (LLN)}}{n \to \infty}
\E \left[ \log \frac{ P_\tsubs{Y}(Y)}{P_\tsubs{Y|X}(Y | X)} \right]
\\& \defeq
-I(X;Y)
,
\end{split}\end{equation}
where $X,Y$ are two random variables distributed according to $P_\tsubs{X}(X) \cdot P_\tsubs{Y|X}(Y|X)$. If mutual information has not been defined, then the last equality may be considered its definition. From the L.L.N. it holds that for any $\epsilon, \delta > 0$ there is $n$ large enough such that with probability at least $1-\epsilon$ (the probability is over $\vr X,\vr Y$):
\begin{equation}\label{eq:94}
\frac{1}{n} \log \frac{ P_\tsubs{Y}(\vr Y)}{P_\tsubs{Y|X}(\vr Y | \vr X)} \leq -I(X;Y)+\delta
\end{equation}
When \eqref{eq:94} holds, then by \eqref{eq:72} the conditional probability of error is bounded by $2^{-n(I(X;Y)-\delta-R)} $ and thus the overall error probability is bounded by the union bound:
\begin{equation}\label{eq:102}
 P_{e} \leq \epsilon + 2^{-n(I(X;Y)-\delta-R)},
\end{equation}
which can be made arbitrarily small if $R < I(X;Y)$, since $\epsilon, \delta$ can be arbitrarily small. This proves $I(X;Y)$ is an achievable rate (by standard definitions, e.g. \cite[\S 7.5]{CoverThomas_InfoTheoryBook}), and the capacity is attained by optimizing over $P_\tsubs{X}(\cdot)$.
\endofproof

The same proof applies to continuous channels, i.e. $P_\tsubs{XY}(\cdot)$ may denote a probability density of continuous variables rather than a probability mass function, and the expression for mutual information directly translates into the continuous expression (difference between differential entropies).

\section{The Normal approximation and channel dispersion}
It is simple and instructive to continue the argument above and develop the well known Normal approximation for the gap from capacity required for a certain error probability. This is a well known result attributed to Strassen and tightened by Polyanskiy~\etal~\cite[Thm.45]{Polyanskiy2010}. The technique is not new, and the point is to see how it evolves naturally from the previous steps. In the following mathematical details are waved aside.

Recognizing that the term $\frac{ P_\tsubs{Y}(\vr Y)}{P_\tsubs{Y|X}(\vr Y | \vr X)}$ in \eqref{eq:72} is $2^{-i}$ where $i$ is the information density (a function of $\vr X, \vr Y$), rewrite \eqref{eq:72} as $P_{e|xy} \leq 2^{nR -i}$. Replacing the weak LLN argument used in \eqref{eq:82} for this term, by the strong LLN, implies that $\frac{i}{n}$ converges in distribution to the Gaussian distribution $\Normal \left(I, \frac{V}{n} \right)$, where $I$ is the mutual information (the normalized mean of $i$) and $V$ is the dispersion, i.e. $V = \Var \left( \log \frac{ P_\tsubs{Y}(Y)}{P_\tsubs{Y|X}(Y | X)} \right)$.

Ignoring the overhead terms related to this convergence and assuming that indeed $\frac{i}{n} \sim \Normal \left(I, \frac{V}{n} \right)$, then:
\begin{equation}\begin{split}\label{eq:121}
P_e
&=
\E \left[ P_{e|xy} \right]
\\& \leq
1 \cdot \Pr \left\{ P_{e|xy} > 2^{-n \delta} \right\}
\\& \qquad + 2^{-n \delta} \cdot \Pr \left\{ P_{e|xy} \leq 2^{-n \delta} \right\}
\\& \stackrel{\eqref{eq:72}}{\leq}
\Pr \left\{  2^{nR -i} > 2^{-n \delta} \right\} + 2^{-n \delta}
\\& =
\Pr \left\{  \frac{i}{n} < R + \delta \right\} + 2^{-n \delta}
\\& \approx
Q \left( \frac{ I-\delta-R }{\sqrt{V/n}} \right) + 2^{-n \delta}
,
\end{split}\end{equation}
where that $\delta$ is a parameter of choice. Requiring that the RHS equal a desired error probability $\epsilon$, the following rate $R$ is extracted from \eqref{eq:121}:
\begin{equation}\label{eq:153}
R = I - \delta - \sqrt{\frac{V}{n}} Q^{-1} ( \epsilon - 2^{-n \delta} )
.
\end{equation}
By letting $\delta$ decrease slower than $\frac{1}{n}$ but faster than $\frac{1}{\sqrt{n}}$, the term $2^{-n \delta}$ can be made negligible compared to $\epsilon$ while $\delta$ becomes negligible compared to $\sqrt{V/n} Q^{-1} ( \epsilon )$, and since $Q(\cdot)$ is continuous the following well known approximation is obtained:
\begin{equation}\label{eq:153b}
R \approx I - \sqrt{\frac{V}{n}} Q^{-1} ( \epsilon )
.
\end{equation}

\section{Discussion}
The techniques used in the proofs above are all well known. The only new technique is the use of the Markov bound in \eqref{eq:61}. As can be seen in \eqref{eq:121}, the result of \eqref{eq:72} is equivalent to a theorem by Shannon \cite[Thm.1]{ShannonCertain57}\cite[Thm.2]{Polyanskiy2010}. Shannon showed this bound is tight in terms of rate \cite[Thm.2]{ShannonCertain57}. Many results in coding can be obtained from this bound, or from its stronger versions such as Feinstein's Lemma \cite[Thm.1]{Polyanskiy2010} and results by Polyanskiy~\etal~\cite[Lemma 19]{Polyanskiy2010}. Therefore the main technical contribution of this paper is in supplying a short proof of Shannon's theorem \cite[Thm.1]{ShannonCertain57} by \eqref{eq:61}-\eqref{eq:72}.

It may seem surprising that tight bounds can be obtained by Markov inequality. First note, that for analysis of error probability near the channel capacity, the tightness of the bound on $P_{e|xy}$ is not critical. The error probability is typically bounded (as in \eqref{eq:121}) by two components: the probability of the normalized information density $i/n$ to fall below the rate $R$ (the probability of a ``bad empirical channel''), and the remaining error probability when $i/n$ is above the rate (the error probability in a ``good empirical channel''). Near the capacity, the first probability is dominant (as evident from the previous section), and therefore, roughly speaking, any bound on the error that vanishes when $i/n > R + \delta$ is satisfactory. To compare with the methods used in Shannon's proof \cite[Thm.1]{ShannonCertain57}, first use Bayes rule to reformulate the pairwise error condition $P_\tsubs{Y|X}(\vr Y | \vr X_m) \geq P_\tsubs{Y|X}(\vr Y | \vr X)$ as $\frac{P_\tsubs{X|Y}(\vr X_m | \vr Y)}{P_\tsubs{X}(\vr X_m)} \geq 2^i = \frac{P_\tsubs{X|Y}(\vr X | \vr Y)}{P_\tsubs{X}(\vr X)}$. The LHS is the ratio between two probability distributions on $\vr X$. Given two distributions $P,Q$, Shannon's argument is based on the fact that the probability under distribution $P$ of the set $\left\{x: \frac{Q(x)}{P(x)} > t \right\}$ cannot be larger than $\frac{1}{t}$, which is obtained by summing both sides of $P(x) \leq \frac{1}{t} Q(x)$ over the set. Using Markov inequality yields the same bound because $\underset{P}{\E} \left[\frac{Q(x)}{P(x)} \right] = 1$. This explains the fact the two bounding techniques yield the same bound in \eqref{eq:72}. Markov inequality yields a simpler derivation and the information density, and subsequently the mutual information, follow naturally from the bound.

Note that in using the L.L.N. in \eqref{eq:82} the relevant r.v. is required to have a bounded variance.
This assumption appears also in the AEP based proof.


\end{document}